\documentclass[journal,twoside,web]{ieeecolor}
\usepackage{lcsys}
\usepackage{cite}
\usepackage{amsmath,amssymb,amsfonts}
\usepackage{graphicx}
\usepackage{textcomp}
\usepackage{cite}
\usepackage{amsmath,amssymb,amsfonts}
\usepackage{algorithmic}
\usepackage{graphicx}
\usepackage{textcomp}
\usepackage{xcolor}

\newtheorem{assumption}{Assumption}
\newtheorem{theorem}{Theorem}
\newtheorem{corollary}{Corollary}{}
\newtheorem{remark}{Remark}
\newtheorem{definition}{Definition}
\newtheorem{lemma}{Lemma}

\usepackage{amsmath}

\usepackage[ruled,linesnumbered]{algorithm2e}

\def\BibTeX{{\rm B\kern-.05em{\sc i\kern-.025em b}\kern-.08em
    T\kern-.1667em\lower.7ex\hbox{E}\kern-.125emX}}
\begin{document}
\title{Deep Neuro-Adaptive Sliding Mode Controller for Higher-Order Heterogeneous Nonlinear Multi-Agent Teams with Leader}
\author{Khushal Chaudhari, Krishanu Nath and Manas Kumar Bera \IEEEmembership{Member, IEEE} 
\thanks{Manuscript submitted on 12 April 2024
(Corresponding author: Krishanu Nath)
}
\thanks{Khushal Chaudhari, and Manas Kumar Bera are with the Department of Electrical Engineering, National Institute of Technology Rourkela, India (e-mail: 923ee5004@nitrkl.ac.in; beramk@nitrkl.ac.in).}
\thanks{Krishanu Nath is with the Department of Electrical Engineering,
Dr B R Ambedkar National Institute of Technology Jalandhar, India (e-mail: nathk@nitj.ac.in).}
}

\maketitle
\thispagestyle{empty} 
\begin{abstract}
This letter proposes a deep neural network (DNN)-based neuro-adaptive sliding mode control (SMC) strategy for leader-follower tracking in multi-agent systems with higher-order, heterogeneous, nonlinear, and unknown dynamics under external disturbances. The DNN is used to compensate the unknown nonlinear dynamics with higher accuracy than shallow neural networks (NNs) and SMC ensures robust tracking. This framework employs restricted potential functions within a set-theoretic paradigm to ensure system trajectories remain bounded within a compact set, improving robustness against approximation errors and external disturbances. The control scheme is grounded in non-smooth Lyapunov stability theory, with update laws derived for both inner and outer layer network weights of DNN. A numerical example is simulated that showcases the proposed controller's effectiveness, adaptability, and robustness.

\end{abstract}


\section{Introduction}
\label{sec:introduction}


\IEEEPARstart{T}{he} autonomous systems like mobile robots, autonomous underwater vehicles, and unmanned air vehicles (UAVs) operating alone are less advantageous than multi-agent teams. Greater sensor range, improved communication coverage, and resilience to individual agent failure are all made possible by a team. Thus, the fields of monitoring, exploration, and disaster and emergency response \cite{9220149} are of tremendous interest to them. Over the past decades, there has been a lot of interest in the distributed cooperative control of multi-agent systems (MAS), such as the consensus, formation, flocking, or rendezvous tasks of autonomous systems \cite{WANG2016204}.
In the leader-follower consensus problem, a designated leader steers a group of autonomous agents—referred to as followers—toward achieving collective objectives \cite{9557752}.



One of the most crucial issues for the consensus problem of nonlinear multi-agent systems is how to suppress the nonlinearities in the dynamics. Dealing with the nonlinearities is made relatively simple in many results by imposing strict constraints on the nonlinear components in the dynamics of the agents \cite{7864443}, \cite{8543491}, \cite{WANG2020104704}. 
The neural network (NN) is a useful tool for more general nonlinearities because of its universal approximation capability. 
The articles \cite{DAS20102014},  \cite{4760250} and \cite{8734736} consider the scalar nonlinear dynamics for the leader and followers, and NNs with a single hidden layer are used, which may not give the desired function approximation performance. 

The advancement in computational power in recent years has led to a boom in deep learning and its use in a variety of control applications. Due to the superior function approximation capabilities of DNNs over shallow NNs, DNN-based control has grown in popularity \cite{9432951}, \cite{9650517}. 
In  \cite{9432951}, \cite{9650517}, the DNN-based controllers are developed for a single uncertain control-affine nonlinear system to solve tracking problems, which is not straightforward to extend for consensus control of multi-agent systems. 

In the robustness issue, the sliding mode control (SMC) is well known for its ability to render the controlled system insensitive to the class of matched uncertainty \cite{Ferrara1}. In the literature \cite{9220828}, SMC-based controllers are designed to solve the leader-follower multi-agent systems (LF-MAS). However, SMC is not a model-free approach, as it requires knowledge of the system model for controller design.
Recently, the function approximation capability of DNNs has been paired with SMC in \cite{10138671} to deal with the robustness of control-affine nonlinear systems with unknown dynamics. Extending DNN-based SMC from single-agent to multi-agent systems is challenging due to inter-agent coupling, decentralized learning, and the need for coordination using only local information, and how the DNN-based SMC law can be designed for LF-MAS is not covered in these articles.

It should be noted that NNs or DNNs theoretically guarantee function approximation capability only when the function is defined over a compact and solvable subset of the state space \cite{7568975}, \cite{8579528}. A critical challenge in the design of neuro-adaptive control systems lies in ensuring that the closed-loop system trajectories remain within this compact region. If trajectories escape this bounded set, the validity of the NN or DNN approximation deteriorates, potentially compromising the stability and performance of the control system. To mitigate this issue, in \cite{ARABI201937}, the set-theoretic model reference neuro-adaptive control frameworks are proposed for uncertain nonlinear systems. The set-theoretic framework based on generalized restricted potential functions offers key advantages by ensuring that both the synchronization error between the uncertain follower and the leader model, as well as the reconstruction error in learning unknown dynamics, remain bounded within user-defined worst-case limits. This guarantees strict performance under system uncertainties.

Motivated by the above-mentioned discussions, we explore DNN-based SMC of higher-order heterogeneous nonlinear leader-follower consensus problems by guaranteeing that the closed-loop system trajectories remain confined within a predefined compact set. The contributions of this letter are as follows.
\begin{enumerate}
      \item A novel DNN-based neuro-adaptive sliding mode consensus protocol is proposed to solve the leader-follower problem of multi-agent systems with an active leader.
    \item The DNN-based SMC framework is formulated within a set-theoretic approach employing restricted potential functions to ensure that the system trajectories remain confined within a prescribed compact subset of the real coordinate space. 
    
\end{enumerate}
\par Notation used in this letter is as follows: The signum function is denoted by $\text{sgn}(\cdot)$. The trace of a square matrix $\mathcal{A}\in \mathbb{R}^{\mathcal{N}\times \mathcal{N}}$ is denoted by the operator $\text{tr}(\cdot)$. The diagonal matrix is represented by $\text{diag}(\cdot)$.  $\overline{\sigma}(\mathcal{A})$ and $\underline{\sigma}(\mathcal{A})$ represent the maximum and minimum singular values of $\mathcal{A}$.  The symbol $\circ$ is used for the composition of the function, e.g., $(g_1\circ g_2)(x)=g_1(g_2(x))$. The operator $|\cdot|$ represents the absolute value, and the 2-norm of a vector is represented by the $\Vert \cdot \Vert$ and the Frobenius norm of a matrix is represented by the operator $\Vert \cdot \Vert_F$.  The operator $\text{vec}(\cdot)$ transforms the matrix into a column vector, e.g., $\mathcal{A}\in \mathbb{R}^{\mathcal{N}\times \mathcal{M}}$  to $\text{vec}(\mathcal{A})\in \mathbb{R}^{\mathcal{N}\mathcal{M}}$. The operator $\mathcal{K}[\cdot]$ is used to compute Filippov's differential inclusion \cite{1086038}.
\section{Mathematical Preliminaries}
Graph theory is the study of graphs, which are mathematical structures used to model pairwise relations between agents. A graph $\mathcal{G}$ is defined as an ordered pair $\mathcal{G} =(\mathcal{V}, \mathcal{E})$, where:
\begin{itemize}
    \item $\mathcal{V}$ is a set of \text{vertices} or \text{nodes}.
    \item $\mathcal{E} \subseteq \{ \{u, v\} \mid u, v \in V, u \neq v \}$ is a set of \text{edges} that connect pairs of vertices.
\end{itemize}
 In this letter, we consider the directed graph represented by an adjacency matrix $A=[a_{ij}]\in \mathbb{R}^{N \times N}$  where $a_{ij}>0$ if there is a directed edge from node $v_j$ to node $v_i$, that is $(v_j,v_i)\in \mathcal{E}$ and $a_{ij}=0$; otherwise. To ensure that there are no self-loops in the graph, we assume $a_{ii}=0$ for all nodes. The graph Laplacian matrix is $L=D-A$ , where $D$ is the in-degree matrix given by $D=\text{diag}\{d_i\} \in \mathbb{R}^{N \times N}$ with each in-degree $d_i$ calculated as $d_i=\sum_{j=1}^{N} a_{ij}$. In the following discussions, the connectivity of the graph is assumed to be known.
 \begin{lemma}
  Under the assumption that the directed graph is strongly connected \cite{5717920} or a directed spanning tree, assume that the root node can get information from the leader node \cite{5898403}. Defining $B=\text{diag}\{b_i\} \in \mathbb{R}^{N \times N}$, where $N$ represents the number of follower nodes in a multi-agent system,  and $b_i>0$ for at least one node \textit{i}, then $L+B$ is non-singular. Now define  
\begin{align} \label{1}
     q&=[q_1\;q_2\dots q_N]^\top  =(L+B)^{-1}\underline{1}, \\
     P&=\text{diag}\{p_i\} =\text{diag}\{ 1/q_i\},\\
      Q&=P(L+B)+(L+B)^\top  P ,
\end{align}
where $\underline{1}=[1\:1\dots 1]^\top \in \mathbb{R}^N$, then $P$ and $Q$ are symmetric positive definite matrices.
\end{lemma}
The following definition introduces the notion of generalized barrier Lyapunov function.
 \begin{definition}\label{def:1}
Let $\Vert y\Vert_H = \sqrt{y^\top  H y}$ be a weighted Euclidean norm, where $y \in \mathbb{R}^\mathcal{N}$ is a real column vector and $H \in \mathbb{R}^{\mathcal{N} \times \mathcal{N}}_{+}$. We define $\Upsilon(\Vert y\Vert_P)$, $\Upsilon: \mathbb{R}^\mathcal{N} \to \mathbb{R}$, to be a generalized restricted potential function (generalized barrier Lyapunov function) on the set $\Omega_\mu \triangleq \{y: \Vert y\Vert_H \in [0, \mu)\}$, where $\mu \in \mathbb{R}_+$ is an a priori, user-defined constant. The detailed properties and a specific example of such a function can be found in the reference work \cite{ARABI201937}.
\end{definition}
 \section{Multi-agent System Dynamics}
 Consider the higher-order multi-agent system with   $N$  agents where $ N  \geq 1$, every agent has $M$ order states. Let's consider the dynamics for \textit{i}-th  follower agent of a multi-agent system,  
 \begin{align} \label{4}
      \dot{x}_i^m &=x_i^{m+1}, \quad m=1,2,\cdots,M-1\nonumber \\
     \dot{x}_i^{M} &=f_i(x_i,t)+u_i(t)+\omega_i(t),
\end{align}
where $x_i^m \in \mathbb{R}$ with $(m=1,2,\cdots, M)$  is the \textit{m}-th state of \textit{i}-th agent and the state vector of the agent $i$ is given by $x_i=[x_i^1,x_i^2,\cdots,x_i^M]^\top \in\mathbb{R}^M$, $f_i: \mathbb{R}^M \rightarrow \mathbb{R}$ is a continuously differentiable unknown nonlinear function of \textit{i}-th agent. The \textit{ i}-th agent's control input is $u_i \in \mathbb{R}$ and $\omega_i \in \mathbb{R}$ is an unknown bounded external disturbance acting on \textit{i}-th agent, where $i \in {1,2,\cdots, N}$. In compact form, the global dynamics of the multi-agent system \eqref{4} is represented as
 \begin{align} \label{5}
     \dot{x}^m &=x^{m+1}, \quad m=1,2,\cdots,M-1\nonumber \\
     \dot{x}^{M} &=f(x,t)+u(t)+\omega(t),
 \end{align}
where $x^m=[x_1^m, x_2^m,\dots, x_N^m]^\top  \in \mathbb{R}^N$ is the \textit{m}-th state vector of the system, $f=[f_1, f_2,\dots, f_N]^\top$, $u=[u_1, u_2,\dots, u_N]^\top  \in \mathbb{R}^N$ is the control input vector, and $\omega=[\omega_1, \omega_2,\dots, \omega_N]^\top  \in \mathbb{R}^N$ is the global disturbance vector acting on the multi-agent system with the assumption that  $\Vert \omega \Vert \leq \omega_m$.
 The leader dynamics of a multi-agent system is 
 \begin{align} \label{6}
 \dot{x}_0^m &=x_0^{m+1}, \quad m=1,2,\cdots,M-1\nonumber \\
     \dot{x}_0^{M} &=f_0(x_0,t),
 \end{align}
 where  $x_0=[x_0^1, x_0^2,\dots ,x_0^M]^\top  \in \mathbb{R}^M$ the state vector of the leader,  $f_0$ is unknown piecewise continuous in $t$ and locally lipschitz in $x_0$. The local neighbourhood synchronization error for \textit{ i}-th agent  is 
 \begin{align} \label{7}
     e_i^m =\sum_{j \in N_i}{a_{ij}(x_j^m-x_i^m)+b_i(x_0^m-x_i^m)},
 \end{align}
  where $ b_i \in \mathbb{R}_{ \geq 0}$ and $b_i > 0$ for at least one \textit{i}-th agent. 
Now the global synchronization error vector for the system is 
\begin{align} \label{8}
     e^m =-(L+B)(x^m- \underline{x}_0^m),
 \end{align}
where $e^m=[e_1^m, e_2^m,\dots, e_N^m]^\top  \in \mathbb{R}^N$ and $\underline{x}_0^m =\underline{1}x_0^m \in \mathbb{R}^N$.
Differentiating \eqref{8} and substituting \eqref{5} and \eqref{6} gives
\begin{align} \label{9}
\dot{e}^m  &= -(L+B)(\dot{x}^m- \dot{\underline{x}}_0^m)=e^{m+1}\quad m=1,2,\cdots,M-1\nonumber \\
\dot{e}^{M}  &= -(L+B)(\dot{x}^{M}- \dot{\underline{x}}_0^{M})\nonumber \\
&= -(L+B)[f(x)+u(t)+\omega(t)-\underline{f_0}(x_0 , t)],
\end{align}
where $\underline{f_0}(x_0 , t)= \underline{1}f_0(x_0,t).$

Now, the control objective is to design a robust control law such that the synchronization error defined in \eqref{9} converges to zero and all follower agents synchronize perfectly with the leader agent in the presence of unknown nonlinearity and bounded external disturbance.

\section{Controller Design}
Define the sliding mode errors $r_i$ for each follower node $i$ as \cite{5717920,5898403} :
\begin{align} \label{11}
r_i = \sum_{m=1}^{M-1} \lambda_m e_i^m + e_i^M ,
\end{align}
where the design parameter $\lambda_i$ is chosen such that the polynomial $s^{M-1} + \lambda_{M-1}s^{M-2} + \cdots + \lambda_1$ is Hurwitz. The $r_i = 0$ ensures that the sliding mode is achieved and $e_i$ converges to zero. To ensure the system is Hurwitz, the coefficients $\lambda_i$  are selected by expressing the polynomial $s^{M-1} + \lambda_{M-1}s^{M-2} + \cdots + \lambda_1$ as a product of linear terms $(s - \beta_1)(s - \beta_2) \cdots (s - \beta_{M-1})$, where  $\beta_1,\beta_2,\dots,\beta_{M-1}$ are chosen positive real numbers. Define the global sliding mode error as $r = [r_1, r_2, \dots, r_N]^\top $ then $r = \lambda_1 e^1 + \lambda_2 e^2 + \cdots + \lambda_{M-1} e^{M-1} + e^M$. Now, defining $\mathcal{E}_1 = [e^1, e^2, \dots, e^{M-1}]\in \mathbb{R}^{M-1}$, $\mathcal{E}_2 = [e^2, e^3, \dots, e^M]\in\mathbb{R}^{M-1}$, $l = [0, 0, \dots, 0, 1]^\top  \in \mathbb{R}^{M-1}$ and 
$\Lambda = 
\begin{bmatrix}
0 &  \quad & I_{(M-2)\times(M-2)} \\
-\lambda_1 & -\lambda_2 & \cdots & -\lambda_{M-1}
\end{bmatrix}$, where $I_{(M-2)\times(M-2)}\in \mathbb{R}^{(M-2)\times(M-2)}$ represents the identity matrix. Thus, we have
\begin{align} \label{12}
\mathcal{E}_2 = \mathcal{E}_1 \Lambda^\top  + r l^\top .
\end{align}
Since the matrix $\Lambda$ is Hurwitz, given any positive real number $\alpha$, there exists a positive definite matrix $\mathcal{P}_1$, such that the following  Lyapunov equation holds:
\begin{align} \label{13}
\Lambda^\top  \mathcal{P}_1 + \mathcal{P}_1 \Lambda = -\alpha I,
\end{align}
where $I \in \mathbb{R}^{(M-1)\times(M-1)}$ is the identity matrix.

The dynamics of the sliding mode error $r$ is
\begin{align} \label{14}
\dot{r} &= \lambda_1 \dot{e}^1 + \lambda_2 \dot{e}^2 + \cdots + \lambda_{M-1} \dot{e}^{M-1} + \dot{e}^M \nonumber \\
&= \eta -(L+B)[f(x)+u(t)+\omega(t)-\underline{f_0}(x_0 , t)] ,
\end{align}
where,
\begin{align} \label{15}
\eta = \lambda_1 e^2 + \lambda_2 e^3 + \cdots + \lambda_{M-1} e^M = \mathcal{E}_2 \bar{\lambda},
\end{align}
with \( \bar{\lambda} = [\lambda_1, \lambda_2, \dots, \lambda_{M-1}]^\top  \). 

To stabilize the sliding dynamics \eqref{14}, the control $u(t)$ must be designed with the knowledge of $f(x)$. The NN is best suited for approximating unknown nonlinear dynamic \(f(x)\). In this letter, we use DNN for approximating the unknown nonlinear function \(f(x)\) with the help of the universal function approximation property \cite{80265}.

Let's define $\Omega \subset \mathbb{R}^M$ as a compact set and $\mathbb{C}(\Omega)$ as a functional space where $f_i:\Omega \rightarrow \mathbb{R}$ and $x_i\in\Omega$. According to the universal function approximation property, there always exist ideal weights and basis functions such that unknown nonlinear dynamics $f_i\in \mathbb{C}(\Omega)$ for the \textit{i}-th agent can be represented as        
\begin{align} \label{16}
     f_i(x_i,t) = W_i^\top \rho_i(\Phi_i(x_i))+\epsilon_i(x_i),
 \end{align}
where for \textit{i}-th agent, $W_i \in \mathbb{R}^{p\times1}$ is denoted as the unknown ideal output layer weight vector of DNN, $\rho_i:\mathbb{R}^p \rightarrow \mathbb{R}^p$ is denoted as the unknown vector of the ideal activation function for the output layer, $p \in \mathbb{Z}_{>0}$ is denoted as the number of neurons in the output layer, $\epsilon_i(x_i)$ is denoted as the unknown function approximation error, and $\Phi_i:\mathbb{R}^M \rightarrow \mathbb{R}^p$ is denoted as inner layers containing unknown ideal weight matrices and unknown ideal activation functions represented as 
\begin{align} \label{17}
     \Phi_i(x_i)=(V_{i,k}^\top \varphi_{i,k} \circ V_{i,{k-1}}^\top \varphi_{i,k-1}\circ\dots V_{i,1}^\top \varphi_{i,1})(V_{i,0}^\top  x_i),
 \end{align}
 where $k \in \mathbb{Z}_{>0}$ is the number of inner layers, $ V_{i,j} \in \mathbb{R}^{L_j\times L_{j+1}}$ denoted as $\textit{j}$-th the inner layer weight matrix and $j \in \{0,1,2,\dots,k\}$, $\varphi_{i,j}:\mathbb{R}^{L_j}\rightarrow \mathbb{R}^{L_j}$ denoted as $\textit{j}$-th inner layer activation function for  $j \in \{1,2,\dots,k\}$, $L_j\in\mathbb{Z}_{>0}$ for all $j\in \{0,1,2,\dots,k\}$ denote the number of neurons in the $j$-th inner layer with $L_0=M$ and $L_{k+1}=p$.

For the $i$-th agent, the DNN estimate the unknown nonlinear dynamic $f_i$ as
\begin{align} \label{18}
     \hat{f}_i(x_i,t) =\hat{W}_i ^\top  \hat{\rho_i}(\hat{\Phi}_i(x_i)),
 \end{align}
where $\hat{W}_i \in \mathbb{R}^{p \times 1}$ is denoted as estimate of  output layer weight vector, $\hat{\rho}_i:\mathbb{R}^p \rightarrow \mathbb{R}^p$ is denoted as user-selected activation function for output layer and $\hat{\Phi}_i:\mathbb{R}^M \rightarrow \mathbb{R}^p$ is denoted as estimate of inner layers defined as

\begin{align} \label{19}
    \hat{\Phi}_i(x_i)=(\hat{V}_{i,k}^\top \hat{\varphi}_{i,k} \circ \hat{V}^\top _{i,k-1}\hat{\varphi}_{i,{k-1}}\circ\dots\hat{V}^\top _{i,1}\hat{\varphi}_{i,1})(\hat{V}^\top _{i,0}x_i),
 \end{align}
 where $\hat{V}_{i,j} \in \mathbb{R}^{L_j\times L_{j+1}}$ denoted as the estimate of $j$-th inner layer  weight matrix and $j \in \{0,1,2,\dots,k\}$, $\hat{\varphi}_{i,j}:\mathbb{R}^{L_j}\rightarrow \mathbb{R}^{L_j}$ denoted as the user-defined  $j$-th inner layer activation function for $j \in \{1,2,\dots,k\}$. The estimation error for the output layer weight vector and the inner layer weight matrix is defined as   $\tilde{W}_i =W_i-\hat{W}_i$ and $\tilde{V}_{i,j} =V_{i,j}-\hat{V}_{i,j}.$
 So for the overall multi-agent system, the global unknown nonlinearity $f$ is represented as
 \begin{align} \label{22}
     f(x,t) = W^\top \rho(\Phi(x))+\epsilon(x),
 \end{align}
 where 
 $f=[f_1, f_2, \dots, f_N]^\top $, $W ^ \top =\text{diag}([W_1^\top , W_2 ^ \top ,\dots, W_N ^ \top ])$, $\rho(\Phi(x))=[\rho_1(\Phi_1(x_1)), \rho_2(\Phi_2(x_2)), \dots, \rho_N(\Phi_N(x_N))]^\top $ and $\epsilon(x)=[\epsilon_1(x_1), \epsilon_2(x_2), \dots, \epsilon_N(x_N)]^\top $.
The estimate for the global unknown nonlinear dynamic $f$ is represented as
\begin{align} \label{23}
     \hat{f}(x,t) =\hat{W}^\top  \hat{\rho}(\hat{\Phi}(x)),
 \end{align}
 where $\hat{f}=[\hat{f}_1,\hat{f}_2,\dots, \hat{f}_N]^\top $, $\hat{W}^\top =\text{diag}([\hat{W}_1^\top , \hat{W}_2^\top ,\dots,\hat{W}_N^\top ]$, $\hat{\rho}(\hat{\Phi}(x))=[\hat{\rho}_1(\hat{\Phi}_1(x_1)),\hat{\rho}_2(\hat{\Phi}_2(x_2)),\dots,\hat{\rho}_N(\hat{\Phi}_N(x_N))]^\top $.
 The global estimation error for the output layer weight matrix and inner layer weight matrix is given by   $\tilde{W} =W-\hat{W}, \quad
    \tilde{V}_j =V_j-\hat{V}_j$,
 where $V_j^\top =\text{diag}([V_{_1,_j}^\top , V_{_2,_j}^\top ,\dots, V_{_N,_j}^\top ])$ and $\hat{V}_j^\top =\text{diag}([\hat{V}_{_1,_j}^\top , \hat{V}_{_2,_j}^\top ,\dots, \hat{V}_{_N,_j}^\top ])$.
The following assumption will be used to prove the stability and to find the gain of the controller.

 \begin{assumption} \label{assump:2}
 For every $i^{\text{th}}$ agent, the outer layer ideal weight matrix $W_i$, the inner layer ideal weight matrix $V_{i,j}$, the ideal activation function $\rho_i$, the user-defined activation function $\hat{\rho}_i$ and the function approximation error $\epsilon_i$ are bounded as $\Vert  W\Vert_F \leq W_m$, $\Vert V_{j}\Vert_F \leq V_m$, $\Vert \rho \Vert \leq \rho_m$, $\Vert \hat{\rho} \Vert \leq \hat{\rho}_m$ and $\Vert \epsilon \Vert \leq \epsilon_m$ in the compact set $\Omega$.
 Moreover, for the leader dynamics, the function $\underline{f_0}(x_0 , t)= \underline{1}f_0(x_0,t)$,   $\forall x_0 \in \Omega_0 \subset \Omega$ is bounded as $\Vert \underline{f_0} \Vert \leq f_m$.
 \end{assumption}
 \begin{remark}\label{rem:1}
While achieving the control objective of stabilization, it must also be ensured that the state trajectories of the multi-agent system remain confined within a compact set $\Omega \subset \mathbb{R}^n$, wherein the DNN approximation retains its validity. 
This objective must be achieved by designing a robust control strategy to compensate the bounded disturbances and approximation error.
Furthermore, the agents' state trajectories are required to achieve synchronization with the leader trajectory specified by 
 \eqref{6}, subject to the condition that the leader's state satisfies $x_0(t) \in \Omega_0$, where $\Omega_0 \subset \Omega$ is a compact set. Additionally, the synchronization error, defined in \eqref{7}, must be constrained within a compact set $\Omega_\mu$, parameterized by a user-defined scalar $\mu$, as formalized in Definition \ref{def:1}. To ensure the boundedness of all relevant signals and preserve the validity of the DNN approximation, the parameter $\mu$ must be selected such that the union of compact sets satisfies $(\Omega_0 \cup \Omega_\mu) \subset \Omega$. 
\end{remark}
\par In order to achieve Remark \ref{rem:1}, we design  the control law for each agent as 
\begin{align} \label{26}
     u_i =& \frac{1}{d_i+b_i}\sum_{m=1}^{M-1}\lambda_me_i^{m+1}+
     \gamma_1 r_i + \gamma_2   \text{sgn}(r_i)\nonumber \\ &-\hat{f}_i(x_i,t)+c_i,
 \end{align}
 where $\gamma_1$, $\gamma_2$ are positive controller gains, and $c_i$ is a corrective signal in order to ensure stability.
   
The control law \eqref{26} for the overall system can be represented as
 \begin{align} \label{27}
     u = (D+B)^{-1}\eta+\gamma_1 r + \gamma_2   \text{sgn}(r)-\hat{f}(x,t)+c,
 \end{align}
where  $\hat{f}(x,t)$ is defined in \eqref{23} and $c=[c_1,c_2,\dots, c_N]^\top $ is overall corrective signal given by 
\begin{align} \label{28}
    c=-(L+B)^{-1}A\hat{W}^\top  \hat{\rho}(\hat{\Phi}(x)).
 \end{align}
 Now substituting \eqref{22}, \eqref{23} , \eqref{27} in      \eqref{14} gives
\begin{align} \label{29}
\dot{r} &= \eta -(L+B)[W^\top \rho(\Phi(x))+\epsilon(x)+(D+B)^{-1}\eta \nonumber\\ &+\gamma_1 r + \gamma_2   \text{sgn}(r)-\hat{W}^\top  \hat{\rho}(\hat{\Phi}(x))+c+\omega(t)-\underline{f_0}(x_0 , t).
 \end{align}

 \begin{figure*}
    \centering
    \includegraphics[width=\linewidth]{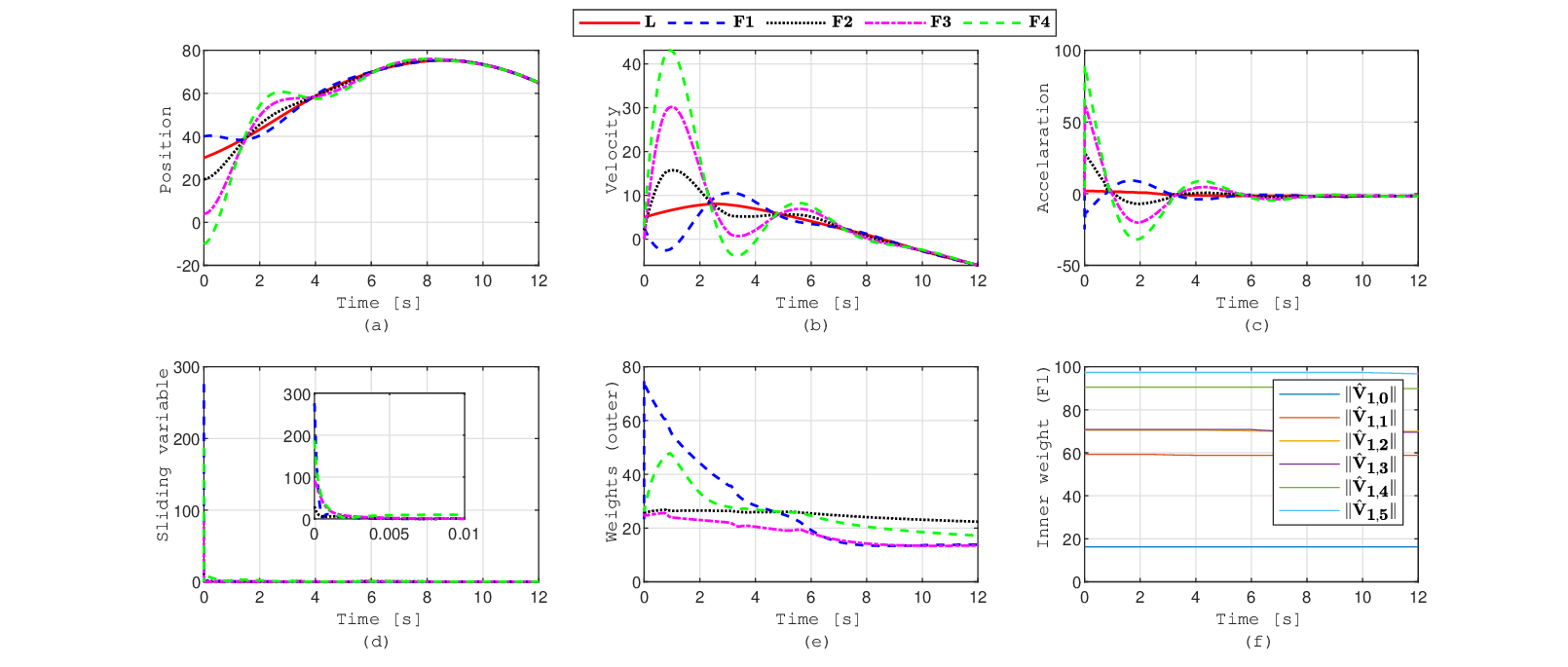}
    \caption{Evolution of (a) position ($x^1$), (b) velocity ($x^2$), (c) acceleration ($x^3$), (d) sliding variable, (e) estimated weights of outer layer, and (f) estimated weights of inner layer of follower 1.}
    \label{fig:enter-label}
\end{figure*}

The outer layer weight update law for each agent is designed as 
\begin{align} \label{30}
     \dot{\hat{W}}_i =-\mathcal{K}_{W_{i}}\hat{\rho}_i(\hat{\Phi}_i(x_i))\Upsilon_d(\sqrt{p_i}r_i)r_ip_i(d_i+b_i)
 \end{align}
where $\mathcal{K}_{W_{i}} \in \mathbb{R}^{p \times p}$ is a positive definite tuning gain matrix.  Now, the global outer layer weight update tuning law for the system is designed as
\begin{align} \label{31}
     \dot{\hat{W}} =-\mathcal{K}_W\hat{\rho}(\hat{\Phi}(x))\Upsilon_d(\Vert r \Vert_P)r^\top  P(D+B).
 \end{align}
 We used a real-time modular DNN approach \cite{9432951} to design the inner layer weights update law. The local inner layer weights update law is designed as      
\begin{align} \label{32}
     \dot{\hat{V}}_{i,j} =&-s_{i,j}(t)v_{i,j}(r_i,t)\mathbf{1}_{\{{\underline{\hat{V}}_{j} \le \Vert \hat{V}_{i,j}\Vert_F \le \overline{\hat{V}}_{j}} \}}\nonumber \\  &\Upsilon_d(\sqrt{p_i}r_i)(d_i+b_i),
 \end{align}
where $s_{i,j}\in \{0,1\}$ is a switching signal used to indicate which inner layer weight update law is active, and this switching approach, which involves layer-by-layer weight updates, helps to mitigate overfitting and ensures bounded adaptation of the DNN parameters. The $\mathbf{1}_{\{\cdot\}}$ is an indicator function takes the value $1$ when the input belongs to the set ${\{\cdot\}}$ and $0$ otherwise, ensuring that the DNN weights remain within a safe and meaningful range during adaptation. Here, the $\underline{\hat{V}}_{j}, \overline{\hat{V}}_{j} \in \mathbb{R}$ are the user-defined lower and upper bounds of $\Vert \hat{V}_{i,j}\Vert_F$, respectively, such that  $\overline{\hat{V}}_{j} \leq V_m$ where $V_m$ is defined in  Assumption \ref{assump:2}. The $v_{i,j} \in \mathbb{R}^{L_j \times L_{j+1}}$ is a user-defined function that satisfies the following inequality, 
\begin{align} \label{33}
      \Vert v_{i,j}(r_i,t)  \Vert_F \le \psi(\sqrt{p_i}r_i)\sqrt{p_i}r_i,
 \end{align}
 where $\psi :\mathbb{R} \rightarrow \mathbb{R}$ is a positive, invertible and non-decreasing function. Now, the global inner layer weights update law is given by 
\begin{align} \label{34}
     \dot{\hat{V}}_j =-s_{j}(t)v_{j}(r,t)\mathbf{1}_{\{{\underline{\hat{V}}_{j} \le \Vert \hat{V}_{j}\Vert_F \le \overline{\hat{V}}_{j}} \}}\Upsilon_d(\Vert r \Vert_P)(D+B),
 \end{align}
 where $s_j(t)=\text{diag}[s_{1,j}(t), s_{2,j}(t),\dots, s_{N,j}(t)]$ , $v_j(r,t)=\text{diag}[v_{1,j}(r_1,t), v_{2,j}(r_2,t),\dots, v_{N,j}(r_N,t)]$ and \eqref{33} becomes
 \begin{align} \label{35} 
      \Vert v_j(r,t)  \Vert_F \le \psi(\Vert r  \Vert_P)\Vert r  \Vert_P.
 \end{align}
 

 \begin{remark}\label{rem:2}
The DNN is adapted online using real-time state information, eliminating the need for offline training. While this facilitates system-specific learning, it necessitates re-adaptation when the system dynamics or operating conditions vary.
\end{remark}

\begin{theorem}\label{thm:important}
Consider the follower dynamics are defined in \eqref{4} and the leader dynamics are given in \eqref{6} with $x_0(t) \in \Omega_0$. If $\Vert r(0) \Vert_P<\mu$, then the control input defined in \eqref{27}, the output layer weight update law in  \eqref{31}, and the inner layer weight update law in \eqref{34} ensure that $\Vert r \Vert_P<\mu$ for all $t\geq0$ and vanishes to zero, provided that the following  gain condition in the control law  \eqref{27} is satisfied,
\begin{align} \label{36}
\gamma_1>&\frac{\overline{\sigma}(A)}{\overline{\sigma}(L+B)\underline{\sigma}(D+B)}\Vert \bar{\lambda} \Vert\nonumber\\&+\frac{\left(\frac{\overline{\sigma}(A)}{\underline{\sigma}(D+B)} \Vert\Lambda \Vert_F \Vert \bar{\lambda} \Vert+\frac{\overline{\sigma}(P_1)}{\underline{\sigma}(P)}\right)^2}{2\alpha},
\end{align}
\begin{align} \label{37}
     \gamma _2 \geq&(\overline{\sigma}(D+B)/\overline{\sigma}(L+B))(W_m \hat{\rho}_m  +2(k+1)V_m\psi(\mu))  \nonumber \\
     &+W_m \rho_m+\epsilon_m+\omega_m+f_m.
 \end{align}
 Moreover, the DNN weights remain bounded under the proposed update laws.
\end{theorem}
\begin{proof}
    See Appendix I for the proof of Theorem \ref{thm:important}.
 \end{proof}
 The above theorem establishes the existence of sliding motion. Next, we show that the states are confined in a user-defined compact set.
\begin{corollary}\label{cor:transitive}
Consider global sliding mode error defined by $r = \lambda_1 e^1 + \lambda_2 e^2 + \cdots + \lambda_{M-1} e^{M-1} + e^M=\sum_{i=1}^{M-1} \lambda_i e^i + e^M$. If $\|r\|_P < \mu$  with the coefficients satisfy 
 $\sum_{i=1}^{M-1} \lambda_i > 1$ and $\lambda_i > 0 \quad \forall i$, then it follows that $\|e^m\|_P < \mu$ and $\|e^m\|_P$ also vanishes $\quad \forall~m \in \{1, 2, \dots, M\}$ .
\end{corollary}
\begin{proof}
    See Appendix II for the proof of Corollary~\ref{cor:transitive}.
\end{proof}
Finally, this establishes that the proposed control scheme ensures stable tracking for the leader-follower problem in the presence of unknown dynamics and external disturbances.
 \section{Simulations}
\begin{figure}[htp]
    \centering
    \includegraphics[width=5cm]{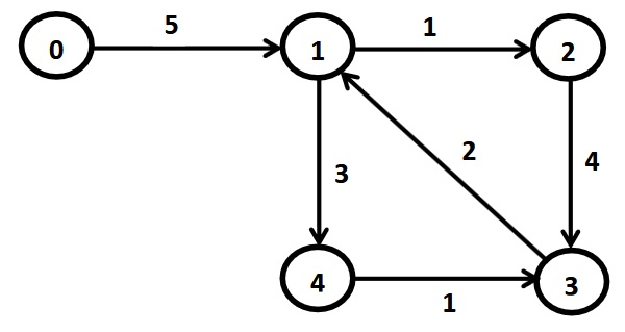}
    \caption{Communication Topology}
    \label{fig:1}
\end{figure}

To demonstrate the applicability and performance of the proposed approach, we provide one illustrative example. Consider the multi-agent system \cite{5717920}  in  Fig. \ref{fig:1}. 
 In Fig. \ref{fig:1}, $0$ represents the leader agent, and $1$ to $4$ represent follower agents. Each agent is represented by a third-order dynamic model, as described in \eqref{4} as
 \begin{align}\label{40}
     \dot{x}_1^{1}&={x}_1^{2}, \quad \dot{x}_1^{2}={x}_1^{3}, \nonumber \\
     \dot{x}_1^{3}&=x_1^2 \sin(x_1^1) + \cos(x_1^3)^2+u_1(t)+\omega_1(t) \nonumber \\
     \dot{x}_2^{1}&={x}_2^{2}, \quad \dot{x}_2^{2}={x}_2^{3}, \nonumber \\ 
     \dot{x}_2^{3} &=x_2^1 + \cos(x_2^2) + (x_2^3)^2 +u_2(t)+\omega_2(t) \nonumber \\
     \dot{x}_3^{1}&={x}_3^{2}, \quad \dot{x}_3^{2}={x}_3^{3}, \nonumber \\ 
     \dot{x}_3^{3} &=x_3^2 + \sin(x_3^3)+u_3(t)+\omega_3(t) \nonumber \\
     \dot{x}_4^{1}&={x}_4^{2}, \quad \dot{x}_4^{2}={x}_4^{3}, \nonumber \\ 
     \dot{x}_4^{3} &=\sin(x_4^1) + (x_4^2)^2 + (x_4^3)^2+u_4(t)+\omega_4(t) , 
 \end{align}
 where bounded disturbances acting on followers are $\omega_1(t)=g_1\text{cos}(t)$, $\omega_2(t)=g_2\text{sin}(t)$, $\omega_3(t)=g_3\text{exp}(-t)$, $\omega_4(t)=g_4\text{sin}(t)\text{cos}(t)$ and $g_1,g_2,g_3$, and $g_4$ are generated randomly in the range $[-5,5]$. The third-order leader dynamics is represented by \eqref{6} as
 \begin{align}
  \dot{x}_0^{1}&={x}_0^{2}, \quad \dot{x}_0^{2}={x}_0^{3}, \nonumber \\ 
  \dot{x}_0^{3}&=- \sin\left((x_0^2)^2\right) - \cos(x_0^3) - e^{-t}. \nonumber 
 \end{align}   
 
To balance computational efficiency and approximation accuracy, through a trial-and-error process, a DNN with six hidden layers is used to approximate unknown nonlinear dynamics. The number of neurons is increased across layers as 10 in the first, 12 in the second, 14 in the third, 15 in the fourth, 18 in the fifth, and 20 in the sixth layer, respectively, and 20 neurons in the output layer. To ensure smooth and differentiable approximations, the hyperbolic tangent (tanh) activation function is utilized uniformly across all neurons. The switching signal of DNN for each agent $s_{ij}(t)$ is designed as
$     s_{i,j}(t) =  \bigg\{ \begin{array}{rcl}
1 
& [2j,\;2(j+1)] \\ 0 & \mbox{else}
\end{array},$
 where $j=\{0,1,2,3,4,5\}$, $i=\{1,2,3,4\}$. Specifically, when $s_{i,j}(t)=1$ the corresponding weight estimate $\hat{V}_{i,j}$ is actively updated based on the adaptive law; conversely, when  $s_{i,j}(t)=0$ the update is suspended, effectively freezing the weight at its current value. 
 The initial values of $\hat{V}_{i,j}$ and $\hat{W}_i$ are randomly selected from a uniform distribution within the range $[-10.5,10.5]$. The lower bound and upper bound of $\Vert\hat{V}_{i,j}\Vert_F$ are taken as $10^{-6}$ and , $250$ respectively. In the inner layer weight update law, the $v_{i,j}$ is selected as
  $   v_{i,j}=\mathcal{K}_{V_{i,j}}\exp(-\frac{r_i^2}{2})\sqrt{p_i}r_i,$ 
 where $j=\{0,1,2,3,4,5\}$, $i=\{1,2,3,4\}$. The learning parameters for the inner layer are taken as $\mathcal{K}_{V_{i,0}}=10\; .\;\textbf{1}_{3\times 10}$, $\mathcal{K}_{V_{i,1}}=10\; .\;\textbf{1}_{10 \times 12}$, $\mathcal{K}_{V_{i,2}}=10\; .\;\textbf{1}_{12 \times 14}$, $\mathcal{K}_{V_{i,3}}=10\; .\;\textbf{1}_{14 \times 15}$, $\mathcal{K}_{V_{i,4}}=10\; .\;\textbf{1}_{15 \times 18}$, $\mathcal{K}_{V_{i,5}}=10\; .\;\textbf{1}_{18 \times 20}$ and the
 learning parameter for the output layer is taken as $\mathcal{K}_{W_i}=10\; .\;\textbf{1}_{20 \times 20}$. Here $\textbf{1}_{n\times m}$ is a $n \times m$ matrix of ones.

The initial value for agents are chosen as $x_0(0)=[30,5,2]$, $x_1(0)=[40,2.6,1]$, $x_2(0)=[20,2.1,2.8]$, $x_3(0)=[4,0.1,-1]$, $x_4(0)=[-10,3,4]$. The $\mu=300$. The tuning gain parameters of the controller are taken as $\gamma_1=500$ and $\gamma_2=0.1$. Design parameters for sliding variables are chosen, $\lambda_1=2$ and  $\lambda_2=1$. 
The response of the multi-agent system is shown in Fig. \ref{fig:enter-label}. The 
Fig. \ref{fig:enter-label} (a)-(c)   effectively demonstrates that all follower agents successfully synchronize with the leader agent, confirming the effectiveness of the proposed control strategy. The Fig. \ref{fig:enter-label} (d) validates that the sliding mode is achieved and $\Vert r \Vert_P$  remains below the specified bound, $\mu$ ensuring the performance constraints are satisfied. The Fig. \ref{fig:enter-label} (e) illustrates the Frobenius norm of the estimated weights in the outer layer $\Vert \hat{W} \Vert_F$  for all five agents, showing consistent adaptation over time. It is observed that the outer layer weights are bounded. In Fig. \ref{fig:enter-label} (f), the Frobenius norm of the estimated weights in the inner layer is shown for Agent 1. Specifically, the estimated weights $\hat{V}_i$ are updated sequentially across different time intervals defined by $s_{ij}(t)$.
These updates reflect the dynamic learning behaviour of the controller and the successful online adaptation of the DNN parameters.

 
\section{Conclusion}
In conclusion, a deep neuro-adaptive sliding mode controller was developed for higher-order heterogeneous nonlinear multi-agent systems with a designated leader. The proposed control framework ensures that all follower trajectories remain bounded within a compact set, thereby preserving the validity of the DNN approximation. The stability of the closed-loop system was rigorously established through a nonsmooth Lyapunov-based analysis. Furthermore, the controller guarantees robust synchronization of the agent trajectories to the leader’s trajectory, despite the presence of unknown external disturbances and nonlinearities. The effectiveness and robustness of the proposed strategy were validated through a comprehensive numerical example. 
Future work may focus on extending the proposed control framework to cases where the network connectivity is not known.

\appendices

\section{Proof of Theorem 1}
Let Lyapunov candidate function as
\begin{align} \label{eq:first}
      \mathcal{V}_L(z,t)=&\frac{1}{2}\Upsilon(\Vert r \Vert_P)+\frac{1}{2}\text{tr}({\mathcal{E}_1\mathcal{P}_1(\mathcal{E}_1)^T})\nonumber \\
      &+\frac{1}{2}\text{tr}(\tilde{W}^T\mathcal{K}_w^{-1}\tilde{W})+\frac{1}{2} \sum_{j=0}^{k} \text{tr}(\tilde{V}_j^T\tilde{V}_j)
 \end{align}
 where $z \in \mathbb{R}^\varkappa$ defined as $z=[r^T\; \mathcal{E}_1^T\;\text{vec}(\tilde{W}^T)\; \text{vec}(\tilde{V}_0)^T\;\text{vec}(\tilde{V}_1)^T\dots\text{vec}(\tilde{V}_k)^T]^T$ and $\varkappa=N(p+1)+(M-1)+\sum_{j=0}^{k}L_jL_{j+1}$.
\par Non-smooth analysis is employed due to the inherent discontinuities introduced by the binary switching signal \(s_{i,j} \in \{0,1\}\), the indicator function \(\mathbf{1}_{\{\cdot\}}\) and the discontinuous components present in the control input itself.
Let $\xi\in \mathbb{R}$ be a Filippov solution to the differential inclusion
\begin{align} \label{eq:second}
    \dot{\xi} \in\mathcal{K}[g](\xi, t),
\end{align}
where $\xi(t) = z(t)$. The calculus of $\mathcal{K}[ \cdot ]$ is used to compute Filippov’s differential inclusion as defined in \cite{1086038}, and $g: \mathbb{R} \times \mathbb{R}_{\geq 0} \to \mathbb{R}$ is defined as $g(\xi, t) = [ \dot{r}^T \; \dot{\mathcal{E}}_1^T \; \text{vec}(\dot{\tilde{W}})^T  \; \text{vec}(\dot{\tilde{V}}_0)^T \dots \text{vec}(\dot{\tilde{V}}_k)^T ]^T$. The generalized time-derivative of $\mathcal{V}_L$ along the Filippov trajectories of $\dot{\xi} = z(\xi, t)$ is defined by
\begin{align} \label{eq:third}
    \dot{\tilde{\mathcal{V}}}_L(\xi, t) = \bigcap_{{\zeta  \in \partial \mathcal{V}_L(\xi, t)}}  \zeta ^T \mathcal{K}[h](\xi, t),
\end{align}
where $\partial \mathcal{V}_L(\xi, t)$ denotes Clarke’s generalized gradient of $\mathcal{V}_L(\xi, t)$ \cite{317122}. Since $\mathcal{V}_L(\xi, t)$ is continuously differentiable in $\xi$, we have $\partial \mathcal{V}_L(\xi, t) = \{ \nabla \mathcal{V}_L(\xi, t) \}$, where $\nabla$ denotes the gradient operator. Also, the time derivative of $\mathcal{V}_L$ exists almost everywhere, i.e., $\dot{\mathcal{V}}_L(\xi, t)  \in \dot{\tilde{\mathcal{V}}}_L(\xi, t) \text{ for almost all } t \in \mathbb{R}_{\geq 0}$. Performing the generalized time derivative of \eqref{eq:first} gives
\begin{align} \label{eq:fourth}
      \dot{\tilde{\mathcal{V}}}_L \subseteq &\frac{1}{2} \frac{d}{dt}\Upsilon(\Vert r \Vert_P)+\text{tr}({\dot{\mathcal{E}}_1 \mathcal{P}_1(\mathcal{E}_1)^T})\nonumber \\ &+\text{tr}(\tilde{W}^T\mathcal{K}_w^{-1} \dot{\tilde{W}})+ \sum_{j=0}^{k} \text{tr}(\tilde{V}_j^T\dot{\tilde{V}}_j)
 \end{align}
 So,
 \begin{align} \label{eq:fifth}
      \frac{d}{dt}\Upsilon(\Vert r \Vert_P)&=\frac{d\Upsilon(\Vert r \Vert_P)}{dr^TPr}\frac{dr^TPr}{dt}\nonumber \\
      &=\Upsilon_d(\Vert r \Vert_P)2r^TP\dot{r}
 \end{align}
 Substituting (23) and \eqref{eq:fifth} in \eqref{eq:fourth} gives
  \begin{align} \label{eq:sixth}
      \dot{\tilde{\mathcal{V}}}_L \subseteq &\Upsilon_d(\Vert r \Vert_P)r^TP\dot{r}+\text{tr}({\mathcal{E}_2 \mathcal{P}_1(\mathcal{E}_1)^T})\nonumber \\ &-tr(\tilde{W}^T\mathcal{K}_w^{-1} \dot{\hat{W}})- \sum_{j=0}^{k} tr(\tilde{V}_j^T\dot{\hat{V}}_j)
 \end{align}
Substituting (11), (27), (29) and (32) in \eqref{eq:sixth} gives
\begin{align} \label{eq:seventh}
      \dot{\tilde{\mathcal{V}}}_L \subseteq& \Upsilon_d(\Vert r \Vert_P)r^TP[\eta -(L+B)[W^T\rho(\Phi(x))+\epsilon(x)\nonumber\\ &+(D+B)^{-1}\eta +\gamma_1 r + \gamma_2   \mathcal{K}[\text{sgn}(r)]\nonumber
    \\
     &-\hat{W}^T \mathcal{K}[\hat{\rho}(\hat{\Phi}(x))]-(L+B)^{-1}A\hat{W}^T  \mathcal{K}[\hat{\rho}(\hat{\Phi}(x))] \nonumber \\
      &  +\omega(t)-\underline{f_0}(x_0 , t)]\nonumber \\
      &+\text{tr}({\mathcal{E}_1 \Lambda^T \mathcal{P}_1(\mathcal{E}_1)^T}) +\text{tr}({ r l^T \mathcal{P}_1(\mathcal{E}_1)^T})\nonumber
    \\
     & +\text{tr}(\tilde{W}^T \mathcal{K}[\hat{\rho}(\hat{\Phi}(x))]\Upsilon_d(\Vert r \Vert_P)r^TP(D+B)) \nonumber
     \\
      &+\sum_{j=0}^{k} \text{tr}(\mathcal{K}[\tilde{V}_j^Ts_j(t)v_j(r,t)\mathbf{1}_{\{{\underline{\hat{V}}_j \le \Vert \hat{V}_j  \Vert_F \le \overline{\hat{V}}_j} \}}]\nonumber
     \\
      &\Upsilon_d(\Vert r \Vert_P)(D+B))
 \end{align}
With help of trace property $\text{tr}(ab^T)=b^Ta$ if  $a,b\in\mathbb{R}^N$, \eqref{eq:seventh}  further solving and rearranging   
\begin{align} \label{eq:eighth}
      \dot{\tilde{\mathcal{V}}}_L \subseteq& \Upsilon_d(\Vert r \Vert_P)r^TP(I-(L+B)(D+B)^{-1})\eta \nonumber\\ &-\Upsilon_d(\Vert r \Vert_P)r^TP(L+B)[W^T\rho(\Phi(x))+\epsilon(x)\nonumber\\ &+ \gamma_1 r + \gamma_2   \mathcal{K}[\text{sgn}(r)]+\omega(t)-\underline{f_0}(x_0 , t)]\nonumber \\
      &+\text{tr}({\mathcal{E}_1 \Lambda^T \mathcal{P}_1(\mathcal{E}_1)^T}) +\text{tr}({ r l^T \mathcal{P}_1(\mathcal{E}_1)^T})\nonumber
    \\
     &+\Upsilon_d(\Vert r \Vert_P)r^TP(L+B)\hat{W}^T \mathcal{K}[\hat{\rho}(\hat{\Phi}(x))]\nonumber
    \\
     &+\Upsilon_d(\Vert r \Vert_P)r^TPA\hat{W}^T  \mathcal{K}[\hat{\rho}(\hat{\Phi}(x))]  \nonumber
    \\
     & +\Upsilon_d(\Vert r \Vert_P)r^TP(D+B)\tilde{W}^T \mathcal{K}[\hat{\rho}(\hat{\Phi}(x))] \nonumber
     \\
      &+\sum_{j=0}^{k} \text{tr}(\mathcal{K}[\tilde{V}_j^Ts_j(t)v_j(r,t)\mathbf{1}_{\{{\underline{\hat{V}}_j \le \Vert \hat{V}_j  \Vert_F \le \overline{\hat{V}}_j} \}}]\nonumber
     \\
      &\Upsilon_d(\Vert r \Vert_P)(D+B)).
 \end{align}
 With help of trace property $\text{tr}(A+A^T)=2\text{tr}(A)$ if  $A$ is square matrix, using $I-(L+B)(D+B)^{-1}=A(D+B)^{-1}$ and $L=D-A$, further solving \eqref{eq:eighth} and rearranging 
 \begin{align} \label{eq:ninth}
       \dot{\tilde{\mathcal{V}}}_L \subseteq& \Upsilon_d(\Vert r \Vert_P)r^TPA(D+B)^{-1}\eta \nonumber\\ &-\Upsilon_d(\Vert r \Vert_P)r^TP(L+B)[W^T\rho(\Phi(x))+\epsilon(x)\nonumber\\ &+ \gamma_1 r + \gamma_2   \mathcal{K}[\text{sgn}(r)]+\omega(t)-\underline{f_0}(x_0 , t)]\nonumber \\
      &+\frac{1}{2} \text{tr}({\mathcal{E}_1 (\Lambda^T \mathcal{P}_1+\mathcal{P}_1\Lambda)(\mathcal{E}_1)^T}) +\text{tr}({ r l^T \mathcal{P}_1(\mathcal{E}_1)^T})\nonumber
    \\
     &+\Upsilon_d(\Vert r \Vert_P)r^TP(D+B)\hat{W}^T \mathcal{K}[\hat{\rho}(\hat{\Phi}(x))]\nonumber
    \\
     &-\Upsilon_d(\Vert r \Vert_P)r^TPA\hat{W}^T \mathcal{K}[\hat{\rho}(\hat{\Phi}(x))]\nonumber
    \\
     &+\Upsilon_d(\Vert r \Vert_P)r^TPA\hat{W}^T  \mathcal{K}[\hat{\rho}(\hat{\Phi}(x))]  \nonumber
    \\
     & +\Upsilon_d(\Vert r \Vert_P)r^TP(D+B) {W}^T \mathcal{K}[\hat{\rho}(\hat{\Phi}(x))] \nonumber
     \\
      &-\Upsilon_d(\Vert r \Vert_P)r^TP(D+B)\hat{W}^T \mathcal{K}[\hat{\rho}(\hat{\Phi}(x))] \nonumber
     \\
      &+\sum_{j=0}^{k} \text{tr}(\mathcal{K}[\tilde{V}_j^Ts_j(t)v_j(r,t)\mathbf{1}_{\{{\underline{\hat{V}}_j \le \Vert \hat{V}_j  \Vert_F \le \overline{\hat{V}}_j} \}}]\nonumber
     \\
      &\Upsilon_d(\Vert r \Vert_P)(D+B)).
 \end{align}
 
 Now substituting (12) and further solving gives 
 \begin{align} \label{eq:tenth}
      \dot{\tilde{\mathcal{V}}}_L \subseteq& \Upsilon_d(\Vert r \Vert_P)r^TPA(D+B)^{-1}\mathcal{E}_1\Lambda^T\bar{\lambda} \nonumber\\ &+ \Upsilon_d(\Vert r \Vert_P)r^TPA(D+B)^{-1}r l^T\bar{\lambda} \nonumber\\ &-\Upsilon_d(\Vert r \Vert_P)r^TP(L+B)[W^T\rho(\Phi(x))+\epsilon(x)\nonumber\\ &+ \gamma_1 r + \gamma_2   \mathcal{K}[\text{sgn}(r)]+\omega(t)-\underline{f_0}(x_0 , t)]\nonumber \\
      &-\frac{\alpha}{2} \text{tr}({\mathcal{E}_1 (\mathcal{E}_1)^T}) +\text{tr}({ r l^T \mathcal{P}_1(\mathcal{E}_1)^T})\nonumber
    \\
     & +\Upsilon_d(\Vert r \Vert_P)r^TP(D+B) {W}^T \mathcal{K}[\hat{\rho}(\hat{\Phi}(x))] \nonumber
     \\
      &+\sum_{j=0}^{k} \text{tr}(\mathcal{K}[\tilde{V}_j^Ts_j(t)v_j(r,t)\mathbf{1}_{\{{\underline{\hat{V}}_j \le \Vert \hat{V}_j  \Vert_F \le \overline{\hat{V}}_j} \}}]\nonumber
     \\
      &\Upsilon_d(\Vert r \Vert_P)(D+B)).
 \end{align}
Using the definition of the calculus \(\mathcal{K}[.]\), $e^TP\mathcal{K}[\text{sgn}(r)]=\Vert r \Vert_P$. With help of (33) and Assumption 1, \eqref{eq:tenth} can be upper bounded as
\begin{align} \label{eq:eleventh}
      \dot{{\mathcal{V}}}_L \leq &\Upsilon_d(\Vert r \Vert_P)\frac{\overline{\sigma}(A)}{\underline{\sigma}(D+B)}\Vert r \Vert_P \Vert \mathcal{E}_1\Vert_F \Vert\Lambda \Vert_F \Vert \bar{\lambda} \Vert  \nonumber \\ &+\Upsilon_d(\Vert r \Vert_P)\frac{\overline{\sigma}(A)}{\underline{\sigma}(D+B)}\Vert r \Vert_P^2 \Vert l\Vert_F \Vert \bar{\lambda} \Vert  \nonumber \\&+ \Upsilon_d(\Vert r \Vert_P)\overline{\sigma}(L+B)\Vert r \Vert_P (W_m \rho_m +\epsilon_m - \gamma _2\nonumber \\
      &+\omega_m+f_m)- \Upsilon_d(\Vert r \Vert_P)\overline{\sigma}(L+B)\gamma_1 \Vert r \Vert_P^2\nonumber \\
      &-\frac{\alpha}{2}\Vert \mathcal{E}_1\Vert_F^2+\frac{\overline{\sigma}(\mathcal{P}_1)}{\underline{\sigma}(P)}\Vert r \Vert_P \Vert l\Vert_F \Vert E_1 \Vert\nonumber \\
      &+\Upsilon_d(\Vert r \Vert_P)\overline{\sigma}(D+B)\Vert r \Vert_P(W_m \hat{\rho}_m) \nonumber \\
      &+\Upsilon_d(\Vert r \Vert_P)\overline{\sigma}(D+B)\Vert r \Vert_P(2(k+1)V_m\psi(\Vert r  \Vert_P)), 
 \end{align}
 where $\overline{\sigma}(L+B)$ is the maximum singular value of $L+B$, $\overline{\sigma}(A)$ is the maximum singular value of $A$, $\overline{\sigma}(\mathcal{P}_1)$ is the maximum singular value of $P_1$, $\underline{\sigma}(D+B)$ is the minimum singular value of, $D+B$ and $\underline{\sigma}(P)$ is the minimum singular value of $P$. As $\Vert l\Vert_F=1$, further solving and  rearranging gives
 \begin{align} \label{eq:twelfth}
      \dot{{\mathcal{V}}}_L \leq &(\frac{\overline{\sigma}(A)}{\underline{\sigma}(D+B)} \Vert\Lambda \Vert_F \Vert \bar{\lambda} \Vert+\frac{\overline{\sigma}(\mathcal{P}_1)}{\underline{\sigma}(P)})\Vert r \Vert_P  \Vert \mathcal{E}_1 \Vert  \nonumber \\ &-(\overline{\sigma}(L+B)\gamma_1-\frac{\overline{\sigma}(A)}{\underline{\sigma}(D+B)}\Vert \bar{\lambda} \Vert)\Vert r \Vert_P^2    \nonumber \\&-\frac{\alpha}{2}\Vert \mathcal{E}_1\Vert_F^2- ((\overline{\sigma}(L+B) (\gamma _2-W_m \rho_m -\epsilon_m\nonumber \\
      &-\omega_m-f_m))-\overline{\sigma}(D+B)(W_m \hat{\rho}_m) \nonumber \\
      &-\overline{\sigma}(D+B)(2(k+1)V_m\psi(\Vert r  \Vert_P)))\Vert r \Vert_P.
 \end{align}
 To make $\gamma _2 >(\overline{\sigma}(D+B)/\overline{\sigma}(L+B))(W_m \hat{\rho}_m  +2(k+1)V_m\psi(\Vert r  \Vert_P))+W_m \rho_m+\epsilon_m+\omega_m+f_m$, choosing  $\gamma _2 \geq(\overline{\sigma}(D+B)/\overline{\sigma}(L+B))(W_m \hat{\rho}_m  +2(k+1)V_m\psi(\mu))+W_m \rho_m+\epsilon_m+\omega_m+f_m$. Solving gets $\Vert r  \Vert_P<\mu$. So, \eqref{eq:twelfth} can be further upper bounded as
 \begin{align} \label{eq:thirteenth}
    \dot{\mathcal{V}}_L \leq 
    -\begin{bmatrix}
    \Vert r \Vert_p & \Vert \mathcal{E}_1 \Vert_F
    \end{bmatrix}
    M
    \begin{bmatrix}
    \Vert r \Vert_p \\
    \Vert \mathcal{E}_1 \Vert_F
    \end{bmatrix}, \quad \forall r \in \Omega_\mu
\end{align}
 where
 $\Omega_\mu=\{ r  \in \mathbb{R}^N :\Vert r  \Vert_P<\mu\}$
 and $M=\begin{bmatrix}
m_1 & m_2 \\
m_2 & m_3
\end{bmatrix}$
Now $m_1$, $m_2$, $m_3$ are
\begin{align} \label{eq:fourteenth}
       m_1=\overline{\sigma}(L+B)\gamma_1-\frac{\overline{\sigma}(A)}{\underline{\sigma}(D+B)}\Vert \bar{\lambda} \Vert
\end{align}
\begin{align} \label{eq:fifteenth}
       m_2=-\frac{\frac{\overline{\sigma}(A)}{\underline{\sigma}(D+B)} \Vert\Lambda \Vert_F \Vert \bar{\lambda} \Vert+\frac{\overline{\sigma}(\mathcal{P}_1)}{\underline{\sigma}(P)}}{2}
\end{align}
\begin{align} \label{eq:sixteenth}
       m_3=\frac{\alpha}{2}.
\end{align}
To make $M$  positive definite, selecting $\gamma_1> \frac{\overline{\sigma}(A)}{\overline{\sigma}(L+B)\underline{\sigma}(D+B)}\Vert \bar{\lambda} \Vert +\frac{(\frac{\overline{\sigma}(A)}{\underline{\sigma}(D+B)} \Vert\Lambda \Vert_F \Vert \bar{\lambda} \Vert+\frac{\overline{\sigma}(\mathcal{P}_1)}{\underline{\sigma}(P)})^2}{2\beta}$.
 From \eqref{eq:first} and \eqref{eq:thirteenth}, $\mathcal{V}_L$ is bounded, which implies $r\;, \mathcal{E}_1\;\tilde{W},\;\tilde{V}_j$ are bounded.
 If $r$ bounded, then $v_j$ is bounded from (33) and hence $\dot{\hat{V}}_j$ is bounded. If $\tilde{W}$ is bounded, then $\hat{W}$ is bounded as $W$ is bounded from Assumption 1. If $\tilde{V}_j$ is bounded, then $\hat{V}_j$ is bounded, as $V_j$ is bounded from Assumption 1.  If $x$ and $\hat{V}_j$ are bounded, then $\hat{\Phi}(x)$ is bounded from (16). If $r$ and $\hat{\rho}$ from Assumption 1 are bounded, then $\dot{\hat{W}}$ is bounded from (29). So according to the Lasalle-Yoshizawa theorem extension to nonsmooth systems \cite{6508855},\cite{8360473}, $\begin{bmatrix}
    \Vert r \Vert_p & \Vert \mathcal{E}_1 \Vert_F
    \end{bmatrix}
    M
    \begin{bmatrix}
    \Vert r \Vert_p \\
    \Vert \mathcal{E}_1 \Vert_F
    \end{bmatrix}\rightarrow 0$ implies $\|r\|_P\rightarrow 0$ as $t\rightarrow \infty$.    This completes the proof.
 \hfill \(\blacksquare\)
 \counterwithin{equation}{section}
 
\section{Proof of Corollary 1}
Starting with the expression for \( r \), we write
\begin{align}\label{eq1}
    \|r\|_P = \left\| \sum_{i=1}^{M-1} \lambda_i e^i + e^M \right\|_P
\end{align}
Applying the reverse triangle inequality
\begin{align}\label{eq2}
\|r\|_P \geq \|e^M\|_P - \left\| \sum_{i=1}^{M-1} \lambda_i e^i \right\|_P.
\end{align}
Rearranging terms gives
\begin{align}\label{eq3}
\|e^M\|_P \leq \|r\|_P + \left\| \sum_{i=1}^{M-1} \lambda_i e^i \right\|_P.
\end{align}
Now, applying the triangle inequality
\begin{align}\label{eq4}
\left\| \sum_{i=1}^{M-1} \lambda_i e^i \right\|_P \leq \sum_{i=1}^{M-1} \lambda_i \|e^i\|_P.
\end{align}
Assuming that each agent's error is upper bounded by the maximum agent error \( \|e^i\|_P \leq \|e^M\|_P \), we obtain
\begin{align}\label{eq5}
\left\| \sum_{i=1}^{M-1} \lambda_i e^i \right\|_P \leq \left( \sum_{i=1}^{M-1} \lambda_i \right) \|e^M\|_P.
\end{align}
Substituting this into the earlier bound gives
\begin{align}\label{eq6}
\|e^M\|_P \leq \|r\|_P + \left( \sum_{i=1}^{M-1} \lambda_i \right) \|e^M\|_P.
\end{align}
Rearranging terms
\begin{align}\label{eq7}
\|e^M\|_P \left( 1 - \sum_{i=1}^{M-1} \lambda_i \right) \leq \|r\|_P.
\end{align}
Since \( \sum_{i=1}^{M-1} \lambda_i > 1 \) and $\|r\|_P<\mu$  gives
\begin{align}\label{eq8}
\|e^M\|_P \left( 1 - \sum_{i=1}^{M-1} \lambda_i \right) \leq\mu
\end{align}
Now, further solving 
\begin{align}\label{eq9}
\|e^M\|_P \leq \frac{\mu}{\left( 1 - \sum_{i=1}^{M-1} \lambda_i \right)} 
\end{align}
and this will always ensure 
\begin{align}\label{eq10}
\|e^M\|_P < \mu.
\end{align}
 and hence
 \begin{align}\label{eq11}
\|e^m\|_P < \mu \quad \forall m.
\end{align} 
Since we have shown that $\|r\|_P\rightarrow 0$ in Appendix I, it follows that $\|e^m\|_P\rightarrow 0$ as $t\rightarrow \infty \quad \forall m$.
Thus, the proof is completed.
 \hfill \(\blacksquare\)

\bibliography{Main_merged}
\bibliographystyle{ieeetr}

\end{document}